\documentclass[12pt]{amsart}
\usepackage[foot]{amsaddr}
\usepackage[T1]{fontenc}
\usepackage[utf8]{inputenc}
\usepackage{graphicx} 
\usepackage{color}
\usepackage{xcolor}
\usepackage{marginnote}
\usepackage{lscape}
\usepackage{amsmath}
\usepackage{amsfonts}
\usepackage{amssymb}

\begin{document}

\title{Evaluating virtual-control-augmented trials for reproducing treatment effects from original RCTs}

\author{Alex Fernandes}
\email{alex.fernandes@u-paris.fr}

\author{Raphaël Porcher}
\email{raphael.porcher@u-paris.fr}

\author{Viet-Thi Tran} 
\email{thi.tran-viet@aphp.fr}

\author{François Petit}
\email{francois.petit@inserm.fr}

\address{\hspace{-0.25cm} *Centre d’Épidémiologie Clinique, Hôpital Hôtel-Dieu, Assistance Publique-Hôpitaux de Paris (AP-HP), Université Paris Cité and Université Sorbonne Paris Nord, Inserm, INRAE, Centre for Research in Epidemiology and Statistics (CRESS), 75004, Paris, France}
\address{\quad}
\address{\hspace{-0.25cm} $\dagger$ Université Paris Cité and Université Sorbonne Paris Nord, Inserm, INRAE, Centre for Research in Epidemiology and Statistics (CRESS), 75004, Paris, France}
\address{\quad}

\thanks{F.P. was supported by the French Agence
Nationale de la Recherche through the project reference ANR-22-CPJ1-0047-01.}

\date{\today}

\renewcommand{\shortauthors}{A. Fernandes, R. Porcher, V-T. Tran, F. Petit}
\renewcommand{\shorttitle}{Virtual-control for reproducing treatment effects}

\newcommand{\Var}{\mathrm{Var}}
\newcommand{\MSE}{\mathrm{MSE}}
\newcommand{\E}{\mathbb{E}}
\newcommand{\R}{\mathbb{R}}
\newcommand{\N}{\mathbb{N}}

\newcommand{\thi}[1]{{\color{blue} #1}}
\newcommand{\alex}[1]{{\color{violet} #1}}
\newcommand{\fr}[1]{{\color{olive} #1}}
\newcommand{\raphael}[1]{{\color{red} #1}}
\newcommand{\trou}{{\color{red} TO FILL }}
\newcommand{\aref}{{\color{red} ADD REF }}

\maketitle

\section*{Abstract}
This study investigates the use of virtual patient data to augment control arms in randomized controlled trials (RCTs). 
Using data from the IST and IST3 trials, we simulated RCTs in which the recruitment in the control arms would stop after a fraction of the initially planned sample size, and would be completed by virtual patients generated by CTGAN and TVAE, two AI algorithms trained on the recruited control patients. 
In IST, the absolute risk difference (ARD) on death or dependency at 14 days was $-0.012$ (SE $0.014$).
Completing the control arm by CTGAN-generated virtual patients after the recruitment of $10\%$ and $50\%$ of participants, yielded an ARD of $0.004$ (SE $0.014$) (relative difference $133\%$) and $-0.021$ (SE $0.014$) (relative difference $76\%$), respectively. 
Results were comparable with IST3 or TVAE. This is the first empirical demonstration of the risk of errors and misleading conclusions associated with generating virtual controls solely from trial data.
\section{Introduction}
Randomized controlled trials (RCTs) are the gold standard for evaluating the efficacy and safety therapeutic of interventions. Their results constitute the primary evidence base for regulatory approvals by agencies such as the Food and Drug Administration (FDA) or the European Medicine Agency (EMA), and they play a central role in shaping routine medical practice \cite{Marini2023RCT}. A key  challenge in RCTs is the recruitment of a sufficient sample size to achieve adequate statistical power to detect a clinically meaningful effect. From 20\% to 30\% of RCTs fail to meet their target enrolment, with poor participant recruitment being one of the leading causes of premature trial discontinuation \cite{carlisle_unsuccessful_2015,BSBJKLB17, briel_exploring_2021}.

\bigskip
Generative artificial intelligence methods have been proposed to augment RCTs by adding AI-generated virtual patient data to the data of human participants recruited in the trial \cite{NK24, PK23, PK24, ElKababji_2025}. 
Many situations have been envisioned and we focused here on augmenting the data of RCTs with virtual controls. 
While the performance of generative AI methods for producing virtual patients data is usually assessed through their ability to reproduce the distribution of the characteristics of the training dataset, thereby resulting so-called 'high-fidelity' digital twins \cite{Qian2023Synthcity}, the problem in RCTs augmented with virtual patients differs. 
Indeed, here we look at the ability to reproduce the treatment effect that would be obtained if the full trial (relying on physical patients only) had been conducted. 
In that respect, the generative AI model should be able to reproduce the distribution of the characteristics, and the outcome of patients that have not been used for training.

\bigskip

In this study, we aimed to assess the treatment effect estimation abilities of control-augmented RCTs in comparison with standard RCT procedures (i.e., all data come from recruited participants). We used two generative AI algorithms, namely CTGAN and TVAE, on the data from two RCTs, the International Stroke Trial (IST) \cite{IST97} and the third International Stroke Trial (IST3) \cite{IST308}. 
\section{Results}
The IST is a RCT of 19,435 patients with acute ischaemic stroke assessing the safety and efficacy of aspirin and subcutaneous heparin on death or dependency within 14 days. The IST3 is a RCT of 3,035 patients with acute ischaemic stroke assessing the benefits and harms of intravenous thrombolysis with recombinant tissue plasminogen activator within 6 hours on death and dependence (as measured with Oxford Handicap Scale).

\bigskip

\begin{figure}
    \centering
    \includegraphics[scale = 0.40]{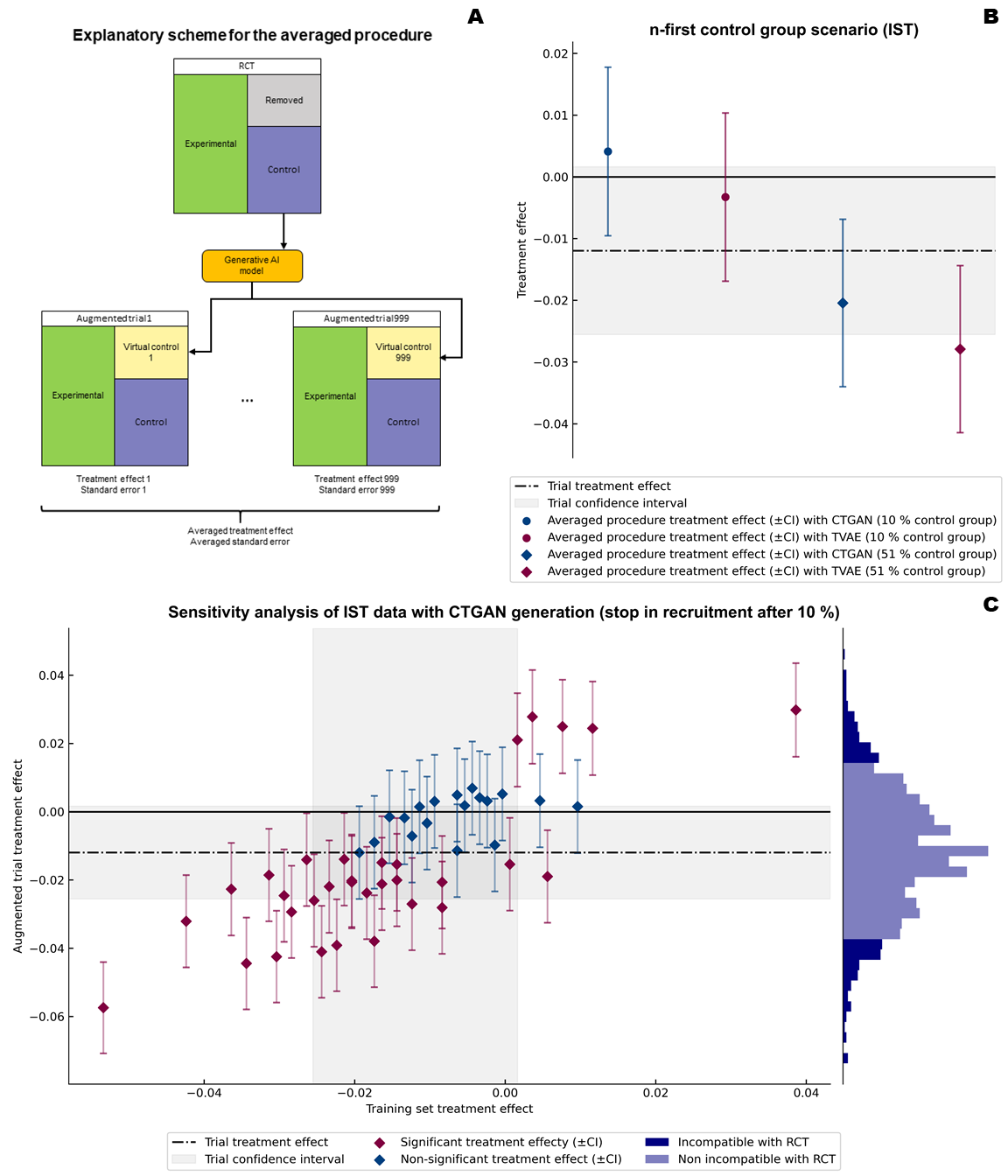}
    \caption{A. Explanatory scheme for the average procedure; B. Treatment effect obtained from the averaged procedure with different architectures and training set size ; C. Treatment effect from augmented trial data relatively to treatment effect of their training set, we ordered the model by the treatment effect obtained from their training set and represented the first one, the last one and one every twenty between them. The left subplot histogram represents the distribution of treatment effect obtained with the averaged procedure.}
    \label{fig:fig}
\end{figure}

In each trial, we estimated the treatment effect obtained if patient recruitment had been stopped after $n$ patients with further patients being generated by artificial intelligence. We trained a model and simulated at first $1$ (one-shot procedure) then $999$ (averaged procedure) augmented trial data where the missing patients data from the original trial were replaced by generated patient data. In the averaged procedure, the $999$ treatment effect and $999$ standard error were averaged (Figure \ref{fig:fig} Panel A). 

This process was repeated on two different training set sizes per trial and using two different architectures CTGAN and TVAE \cite{ctgan, HSSP23}. CTGAN consists of two interlinked neural networks - the generator and the discriminator - that are jointly trained in an adversarial manner while TVAE is also composed of two interlinked neural networks - the encoder and the decoder - that are trained to maximise the Evidence Lower Bound (ELBO) which is a lower bound of the log-likelihood of the data.

\bigskip

In the IST, the absolute risk difference (ARD) on death or dependency at 14 days was $-0.012$ (SE $0.014$).
With the one-shot procedure, the generation of virtual patient data using CTGAN after the recruitment of the $10\%$ and $50\%$ first participants in the IST yielded an ARD of $0.004$ (SE $0.014$) (relative difference $133\%$) and $-0.021$ (SE $0.014$) (relative difference $75\%$), respectively. 
Similar results were found for the other scenarios (see Figure \ref{fig:fig} Panel B and Table \ref{tab:nfirst} from the Annex). 
In the averaged procedure, the generation of virtual patient data using CTGAN after the recruitment of the $10\%$ and $50\%$ first participants in the IST yielded an ARD of $0.004$ (SE $0.014$) (relative difference $133\%$) and an ARD of $-0.020$ (SE $0.014$) (relative difference $67\%$), respectively. 
Similar results were found for the other scenarios (see Figure \ref{fig:fig} Panel B and Table \ref{tab:nfirst} from the Annex).

Whereas both original studies failed to show a significant treatment effect, all trials that were completed after the recruitment of $50\%$ of the control group showed a significant treatment effect.

\bigskip

To mimic the use of generative AI to complete control groups from RCTs to assess the effect related to the variability of the training set, we reproduced the averaged procedure with $1000$ training sets uniformly drawn in the control group patients data from the original trial (as opposed to focusing on the $n$-patients recruited in the control group).
For instance, the scenario using $50\%$ of the patients from the original control arm would correspond to complete with virtual patients a 2:1 RCT.  

Up to $64 \%$ of the estimations yielded a significant (positive or negative) treatment effect while the original data did not conclude the existence of such effect; up to $22\%$ of the treatment effect estimations obtained from the averaged procedure are even incompatible with the randomised controlled trial estimation i.e. their confidence intervals do not overlap (see Figure \ref{fig:fig} Panel C).

To investigate how differences in results from original and augmented trials would be related to the fact that generative AI reproduces the distribution of characteristics of training datasets, we compared the treatment effect estimated with a difference of means on the training data with the treatment effect obtained with the average procedure arising from this training set and found a correlation between those two treatment effects (see Figure \ref{fig:fig} Panel C).
\section{Discussion}

The use of generic purpose generative AI, which only used the data collected from the control arm to generate virtual control patient data provides unreliable estimation of the treatment effect as compared to the actual results of randomized controlled trials. 
In our empirical study, treatment effect estimates of AI-augmented trials could be twice as large as the actual effect measured in the original trial, even changing the sign of the effect in some cases. 

\bigskip 

These results can be explained by the fact that these general purpose generative AI such as VAE and GAN methods, by design, reproduce the distributions observed in the training data. 
Consequently, they yield treatment effect estimates that mirror those in the training set which explains the correlation observed in Figure \ref{fig:fig} Panel C. 
However, recruitment in RCTs may not be homogeneous over time. 
The hypothesis stating that the treatment effect observed in the $n$ first patients is representative of the effect that would be observed with all recruited patients that is sometimes assumed to support trial augmentation may not verified in practice e.g. due to the opening of new centers or to a shifts in patient characteristics during enrolment. 
Even if there would not be any systematic drift related to the randomisation time, the mean outcome in an once undersampled populations do not perfectly reflect the outcome of the sampled population. Indeed, this explains that generating virtual control patients with a training set randomly drawn from the original control patient data we still observed differences between results between the original trials and the augmented trials.

\bigskip

Of note, whenever patients are generated, the nature of the treatment effect evaluated changes. 
The control-augmented trial estimates the treatment effect as learned by the model, rather than the true effect in the target population. 
Notably, the increased statistical power from adding generated patients reinforces confidence in this model-derived estimate, which may not represent the actual treatment effect in the population that would have been recruited in the trial. 

\bigskip

Our experiment has limitations. 
The trials sample sizes were decided by optimizing a minimax criterion—large enough to detect the anticipated treatment effect, yet as small as feasible to respect both ethical and financial constraints.  
This pushes the confidence interval boundary close to zero and any small change in estimated treatment efficacy makes it significant. This explains the large number of significant treatment effect observed with control patient data augmentation.

Second, our simulations were performed on only two trials and different trial data may have generated different results, based on how recruitment was performed in these trials. Third, the sensitivity analysis comes from a bootstrap procedure which does not perfectly reflect the external validity of the observed error in estimation. 

Thirdly, we used data from a trial that did not plan any data-augmentation procedure. In particular, the trial investigators did not try to minimise the variability of the effect along the recruitment phase.

\bigskip

Other approaches such as \cite{TWIN-GPT_2023, RCTTwinGAN} include external information to generate high fidelity virtual patients. 
In particular, the method from \cite{TWIN-GPT_2023} relies on the "world knowledge" contained in GPT type models to incorporate exogenous information and the one of \cite{RCTTwinGAN} relies on GAN conditioned over the external data from electronic health record data.
Nonetheless, these methods have not yet been evaluated for the task studied in this paper, namely completing a control arm. 
In other fields, other approaches to include domain specific information have been developed through the use of mechanistic models \cite{WJXSK22} or bayesian approaches \cite{LV18}. 

\bigskip

Our studies empirically shows that the use of generative AI to generate virtual control patient data provides unreliable estimation of the treatment effect as compared to the actual results of randomized controlled trials when it solely relies on data collected from the $n$-first participants. 
\bibliographystyle{unsrt}
\bibliography{refs}

\newpage

\appendix

\section{Methods}\label{sec:methods}
\subsection*{Data} \label{sec:data}
We used the data from two RCTs, the International Stroke Trial (IST) \cite{IST97} and the third International Stroke Trial (IST3) \cite{IST308}. The IST is a RCT of 19435 patients with suspected acute ischemic stroke assessing the safety and efficacy of aspirin and subcutaneous heparin on death within 14 days and death or dependency at 6 months and the IST3 is a RCT of 3035 patients with acute ischemic stroke and sought to determine whether a wider range of patients may benefit from the administration of intravenous recombinant tissue plasminogen activator (rt-PA) within 3 hours of symptom onset on death and dependency at 6 months. From the raw IST data, a new variable, was constructed to capture whether a patient was dead or dependent at 6 months using the variable \emph{FDEAD} and \emph{FDENNIS}. Countries were categorized into broad geographic areas (Europe, South America, North America, the Middle East, North Asia, South Asia, Africa, and Oceania). Missing data were addressed using multiple imputation chained equations \cite{vanBuuren11}.

\subsection*{Generative models} \label{sec:gen_model}
\subsubsection*{Architectures}
Generative AI, in this communication refers to the use of deep learning algorithm that use a training dataset to generate new data similar to it. The nature of data determines which type of algorithms can be used. Randomized controlled trials data are tabular and contains categorical covariates. 

In our study we used two state-of-the-art latent space based generative models for tabular data with categorical and continuous features \cite{ctgan}
\begin{itemize}
    \item \emph{CTGAN} which is based on Generative Adversarial Networks (GAN) \cite{GAN14}; an architecture consisting of two interlinked neural networks - the generator and the discriminator. These are jointly trained in an adversarial manner: the generator aims to produce realistic synthetic data starting from random noise while the discriminator seeks to differentiate between real data and the generated synthetic samples. The training continues until the discriminator is no longer able to reliably distinguish real data from synthetic one
    \item \emph{TVAE} which is based on Variational Autoencoders (VAE) \cite{Kingma_Welling_2013}; an architecture also composed of two interlinked neural networks - the encoder and the decoder. The encoder maps the input data to a latent space typically of smaller dimension while the decoder maps this latent space back to the input space. The training is performed by maximizing the Evidence Lower Bound (ELBO) which is a lower bound of the log-likelihood of the data. Our implementation used a prior sampling as introduced in \cite{ctgan}.
\end{itemize}
The specificity of CTGAN and TVAE is their preprocessing of the tabular data allowing the aforementioned neural networks to approximate the distribution of the data despite the categorical nature of some of the features. Both architectures are implemented in the Synthetic Data Vault (SDV) library and are two state of the art among the well established recent architectures \cite{HSSP23} adapted to RCTs data.
\subsubsection*{Evaluation of synthetic data}
The evaluation of the quality of the generated data was performed using the SDMetrics general score. This score is the mean of all column score and all column pair score defined as follows. The column score is given by a Kolmogorov-Smirnov test for numerical columns and a total variation distance for categorical columns. The evaluation of the column pair trends is performed with a Pearson coefficient for numerical columns, a normalized contingency table for categorical columns, a normalized contingency table for mixed type columns (the numerical column is discretised into bins). 
\subsection*{Sampling procedure}
In each trial, we estimated the treatment effect obtained by a stop if patient recruitment had been stopped after $n$-patients with further patients being generated by artificial intelligence. We trained different models and distinguish two procedure to incorporate the virtual patient in the treatment effect estimate:
\begin{itemize}
    \item \emph{one-shot procedure} where we simulated $1$ augmented trial data where the missing patients data from the original trial were replaced by generated patient data and analyse this data as if they would have been obtained from a randomised controlled trial;
    \item \emph{averaged procedure} where we simulated $999$ augmented trial data where the missing patients data from the original trial were replaced by generated patient data the arising $999$ treatment effect and $999$ standard error were averaged (Figure \ref{fig:fig} pannel A).
\end{itemize}
The use of the average procedure is justified as, once a model is trained, it is computationally costless to generate more augmented trial data. The averaging of estimates aims to limit the consequence of the sampling inherent to the GAN and VAE architectures and therefore reflect the \emph{learned} treatment effect and standard error.
\subsection*{Statistical analysis} \label{sec:stats}
\newcommand{\train}{\mathrm{train}}
\newcommand{\gen}{\mathrm{gen}}
\newcommand{\os}{\mathrm{os}}
\newcommand{\av}{\mathrm{av}_l}
In a randomized controlled trial context, the difference of means is a unbiased and consistent estimator of the treatment effect. The object of our analysis is to point out the discrepancy between treatment effect estimated via the difference of means using the data of a RCT or using the data of a controlled-augmented trial.

\bigskip 
We start by introducing notations to describe our statistical methodology.
For any positive integer $k$ we write $[k]$ for the set $\{1,\ldots,k\}$. Consider a RCT of size $m$ with covariate space $\mathcal{X}$, primary outcome space $\mathcal{Y} = \{0,1\}$. The data of the RCT is denoted $\mathcal{D}_m \in \mathcal{U}^m$ where $\mathcal{D}_m = (d_1, \ldots, d_m)$ and the $i^{\mathrm{th}}$ patient data is denoted $d_i = (x_i,y_i,a_i)$ where $x_i$ are covariates, $y_i$ her binary primary outcome and $a_i$ her treatment assignment. Let $m_0$ and $m_1$ be respectively the size of the control group and experimental group and denote similarly $\mathcal{D}_{m_0} = (d^0_0, \ldots, d^{0}_{m_0})$ the control group data. 

The dataset $\mathcal{D}_m$ (resp. $\mathcal{D}_{m_0}$) induces an empirical distribution $\mathbf{P}_{m}$ (resp. $\mathbf{P}_{m_0}$) on $\mathcal{U}$. If $(X^*,Y^*,A^*) \sim \mathbf{P}_{m}$ denote by $\mu_a$ the conditional empirical expectation $\E_{m_a}(Y^* \mid A^* = a)$ for $a = 0,1$. The treatment effect $\overline{\tau}$ estimated in the RCT is given by $\overline{\mu_1} - \overline{\mu_0}$ and denote by $\overline{\sigma}$ its variance.

\subsubsection*{Control group data and generative process}

In this subsection, we formalize the generative process of virtual controls

A training set $\mathcal T_n$ of size $n$ is given by $(d^0_{\iota(1)},\ldots,d^0_{\iota(n)}) \in \mathcal{U}^n$ where $\iota : [n] \to [m_0]$ is an injection. Here, the training set  $\mathcal T_n$ will be composed (1) in the case of stop in control recruitment of the data of the $n$ first patients and (2) in the sensitivity analysis case of the data of $n$ patients drawn without replacement from $\mathcal{D}_{m_0}$.

A generative model with $p$ trainable parameters, a latent space of dimension $q$, is specified by the triple
$\bigl(\theta_\bullet, \Gamma_\bullet, \Pi\bigr)$ where $\theta_\bullet : \mathcal{U}^n \to \R^p$ 
corresponds to a training-to-parameters application and $\Gamma_\bullet \colon \mathbb{R}^p \longrightarrow \bigl\{\Gamma_{\theta}\colon \mathbb{N}\times\mathbb{R}^q\to\mathcal{U}^\infty\bigr\}$ maps a set of parameters to a generator function that satisfies
\[
      \Gamma_{\theta}\colon (s,z)\;\longmapsto\;\Gamma_{\theta}(s,z)\;\in\;\mathcal{U}^s,
      \quad
      s\in\mathbb{N}_{\ge1},\;z\in\mathbb{R}^q,
\]
and $ \mathcal{U}^\infty = \bigcup_{s=1}^\infty \mathcal{U}^s$. In particular, $\Gamma_{\theta_{\mathcal{T}_n}}$ is the decoder in a VAE or the generator in a GAN. The sampling prior $\Pi$ is a probability distribution on the latent space $\mathbb{R}^d$.  

Note that we include the training process in our description of a generative model as this training process plays a key role in our subsequent analysis.

\subsubsection*{Estimations}
We will define the estimators $\widehat{\mu}_{s, \os}^{\mathcal{T}_n}$ and $\widehat{\tau}_{s, \av}^{\mathcal{T}_n}$ respectively used to measure the treatment effect in the one-shot and sensitivity analysis procedures.

\bigskip

The randomized controlled trial estimates the treatment effect $\overline{\tau} = \overline{\mu_1} - \overline{\mu_0}$ with its associated confidence interval $[\overline{\tau} - \overline{\delta}, \overline{\tau} + \overline{\delta}]$ where 
\[\overline{\delta} = z_{0.025} \sqrt{\frac{\Var(Y^* \mid A^* =1)}{m_1}+ \frac{\Var(Y^* \mid A^* =0)}{m_0} }\]
and $z_{0.025}$ is the $0.025$ quantile from $\mathcal N (0,1)$.

\bigskip 

For the sake of brevity, we denote by $y^{\mathcal{T}_n, \train}$ the vector of primary outcomes arising from the RCT data and write $Y^{\mathcal{T}_n,\gen}$ for the primary outcomes data from $\Gamma_{\theta_{\mathcal{T}_n}}(s,Z)$  where $s$ is such that $m_0 = n + s$ and $Z \sim \Pi$, namely the virtual patients data.

\bigskip

In the one-shot procedure we denote the mean and variance in the control-augmented trial respectively by
\begin{align*}
\widehat{\mu}_{s, \os}^{\mathcal{T}_n} =& \frac{1}{m_0} \Big( \sum_{i = 1}^{n} y_i^{\mathcal{T}_n,\train} + \sum_{i = 1}^{s} Y_i^{\mathcal{T}_n,\gen} \Big), \\
(\widehat{\sigma}_{s, \os}^{\mathcal{T}_n})^2 =& \frac{1}{m_0} \Big( \sum_{i=1}^n ( y_i^{\mathcal{T}_n,\train}  -\widehat{\mu}_{s, \os}^{\mathcal{T}_n})^2 + \sum_{i = 1}^{s} (Y_i^{\mathcal{T}_n,\gen} - \widehat{\mu}_{s, \os}^{\mathcal{T}_n})^2 \Big).
\end{align*}
We studied the $\overline{\tau}$-estimator given by $\widehat{\tau}_{s, \os}^{\mathcal{T}_n} := \overline{\mu_1} - \widehat{\mu}_{s, \os}^{\mathcal{T}_n}$ and the confidence interval given by $[\widehat{\tau}_{s, \os}^{\mathcal{T}_n} - \widehat{\delta}_{s, \os}^{\mathcal{T}_n} ,\widehat{\tau}_{s, \os}^{\mathcal{T}_n} + \widehat{\delta}_{s, \os}^{\mathcal{T}_n}]$ where 
\[ \widehat{\delta}_{s, \os}^{\mathcal{T}_n} = z_{0.025} \sqrt{\frac{\Var(Y^* \mid A^* = 1)}{m_1}+ \frac{ (\widehat{\sigma}_{s, \os}^{\mathcal{T}_n})^2}{m_0} }.\]

\bigskip

In the averaged procedure, let $Z^1 \ldots Z^l \sim \Pi$ i.i.d and denote by $Y^{\mathcal{T}_n,\gen, j}$ the vectors of primary outcome from $\Gamma_{\theta_{\mathcal{T}_n}}(s, Z^j)$ where $s$ is such that $m_0 = n + s$ the generated primary outcomes of the $j^{th}$. We denote the empirical mean and empirical variance of the $j^{th}$ trial by
\begin{align*}
\widehat{\mu}_{s, j}^{\mathcal{T}_n} =& \frac{1}{m_0} \Big( \sum_{i = 1}^{n} y_i^{\mathcal{T}_n,\train} + \sum_{i = 1}^{s} Y_i^{\mathcal{T}_n,\gen,j} \Big), \\
(\widehat{\sigma}_{s, j}^{\mathcal{T}_n})^2 =& \frac{1}{m_0} \Big( \sum_{i=1}^n ( y_i^{\mathcal{T}_n,\train}  -\widehat{\mu}_{s,j}^{\mathcal{T}_n})^2 + \sum_{i = 1}^{s} (Y_i^{\mathcal{T}_n,\gen,j} - \widehat{\mu}_{s, j}^{\mathcal{T}_n})^2 \Big).
\end{align*}
We studied the $\overline{\tau}$ and $\sigma$ estimators respectively given by
\begin{align*}
&\widehat{\tau}_{s, \av}^{\mathcal{T}_n} = \frac{1}{l} \sum_{j = 1}^l (\overline{\mu_1}-\widehat{\mu}_{s, j}^{\mathcal{T}_n}),& (\sigma_{s,\av}^{\mathcal{T}_n})^2 = \frac{1}{l} \sum_{j = 1}^l (\widehat{\sigma}_{s, j}^{\mathcal{T}_n})^2.
\end{align*}
The associated confidence interval is given by $[\widehat{\tau}_{s, \av}^{\mathcal{T}_n} - \widehat{\delta}_{s, \av}^{\mathcal{T}_n}, \widehat{\tau}_{s, \av}^{\mathcal{T}_n} + \widehat{\delta}_{s, \av}^{\mathcal{T}_n}]$ where $\widehat{\delta}_{s, \av}^{\mathcal{T}_n} = z_{0.025} \sigma_{s,\av}^{\mathcal{T}_n}.$

\bigskip

\noindent We say that the control augmented trial yields a 
\begin{itemize}
    \item significant positive effect if $0 < \widehat{\tau}_{s, \av}^{\mathcal{T}_n} - \widehat{\delta}_{s, \av}^{\mathcal{T}_n}$,
    \item significant positive effect if $0 > \widehat{\tau}_{s, \av}^{\mathcal{T}_n} + \widehat{\delta}_{s, \av}^{\mathcal{T}_n}$,
    \item incompatible decision if $[\widehat{\tau}_{s, \av}^{\mathcal{T}_n} - \widehat{\delta}_{s, \av}^{\mathcal{T}_n}, \widehat{\tau}_{s, \av}^{\mathcal{T}_n} + \widehat{\delta}_{s, \av}^{\mathcal{T}_n}]$ and $[\overline{\tau} - \overline{\delta}, \overline{\tau} + \overline{\delta}]$ are disjoint.
\end{itemize}

\subsubsection*{Training set in the sensitivity analysis}
In order to simulate different patient recruitment scenario we sampled several training sets. For $T_n^1,\ldots,T_n^k \sim \mathbf{P}_{m_0}^{\otimes n}$ i.i.d, we looked at $\widehat{\tau}_{s, \av}^{T}$ as an estimator of $\overline{\tau}$. We also estimated its mean squared error using
\[\widehat{\MSE}(\widehat{\tau}_{s, \av}^{T}) = \frac{1}{k} \sum_{j=1}^k ( \widehat{\tau}_{s, \av}^{T_j} - \overline{\tau})^2.\]

\subsection*{Experiments}
The $n$-first control group patients data augmentation aims to reproduce a stop after the recruitment of the $n$-first control group patients and the completion of this group by virtual patient data. The sensitivity analysis aims to estimate further the impact of a change in case-mix by simulating different recruitment scenarios.  

\bigskip

We realised a total of eight different cases characterized by a triplet given by a reference RCT, a training set size $n$ and a generative AI architecture. We ran both scenarios $n$-first control group patients data augmentation and the sensitivity analysis for each case. We shall describe the exact protocol for each of these scenarios.
\subsubsection*{n-first control group patients data augmentation}

We created a training set $\mathcal{T}_n$ composed of the data from the first $n$ patients included in the control arm of the reference trial. We implemented a hyperparameters tuning with a 5-fold cross-validation gridsearch to avoid overfitting. The grids are summarized in Table \ref{tab:gs}. We selected the set of hyperparameters implemented in SDV that lead to the best SDMetrics general score to train one model. In the case of CTGAN the optimal epochs number hyperparameter was estimated from the generator loss stabilization point.

From the model $999$ control-augmented trial data were created, each one of them composed of a control arm that is the concatenation of the training set and the virtual patients generated by the model and an experimental arm that is the experimental arm from the reference RCT (see Figure \ref{fig:fig} panel A).
We computed the estimators $\widehat{\tau}_{s, \os}^{\mathcal{T}_n}$ and $\widehat{\tau}_{s, \mathrm{av}_{999}}^{\mathcal{T}_n}$ with their confidence interval and represented them in (Figure \ref{fig:fig} panel B).

\subsubsection*{Sensitivity analysis}
In the sensitivity analysis we aim to replicate different recruitment scenarios by sampling $1000$ group of $n$-first recruited patients data among the RCT control group data. 

A gridsearch hyperparameter tuning approach for the $1000$ patients similar to the one we did in the $n$-first control group data augmentation scenario is not computationally tractable.  
Hence, to avoid overfitting, we drawn uniformly from the control group data three training sets of size $n$ and used a gridsearch approach with a 5-fold cross-validation hyperparameters tuning of the different models on each of the three training sets. Then, we selected averaged SDMetrics general score of each hyperparameter combination over the training set and chose the hyperparameters leading to the best averaged score. The grids considered are summarized in Table \ref{tab:gs}. In the case of CTGAN the optimal epochs number hyperparameter was estimated from the generator loss stabilization point.

\bigskip

We drawn uniformly from the control group data $1000$ training sets of size $n$ which lead to 1000 different models. Every model generated 999 augmented trial data that are composed of a control arm that is the concatenation of the training set and the virtual patients generated by the model and an experimental arm that is the experimental arm from the reference RCT (see Figure \ref{fig:fig} panel A). 

\bigskip 

We computed the estimator  $\widehat{\tau}_{s, \mathrm{av}_{999}}^{j}$ for every $j \in \{ 1, \ldots, 1000\}$ with their confidence interval and represented $5\%$ of them in the panel C from Figure \ref{fig:fig}. We also computed the number of significant negative effect, significant positive effect, incompatible effect and reported it in Table \ref{tab:sensitivity}. We also reported the mean squared error $\widehat{\MSE}(\widehat{\tau}_{s, \mathrm{av}_{999}})$.

\newpage
\section{Tables} \label{sec:tables}

\begin{table}[ht]
\center
\begin{tabular}{|l|c|c|c|c|} 
\hline
\multicolumn{5}{|c|}{n-first control group patients data augmentation}\\
\hline
Model & \multicolumn{2}{|c|}{CTGAN} & \multicolumn{2}{|c|}{TVAE} \\
\hline
\hline
Trial & \multicolumn{4}{|c|}{IST} \\
\hline
Training size & 1000 & 5000 & 1000 & 5000 \\ 
\hline
Trial treatment effect & \multicolumn{4}{|c|}{-0.012} \\
Trial treatment effect standard error & \multicolumn{4}{|c|}{$ 0.014$} \\
\hline
One shot procedure treatment effect & 0.004 & -0.021 & -0.005 & -0.025 \\
One shot procedure standard error & $ 0.014$ & $ 0.014$ & $ 0.014$ & $ 0.014$ \\
Averaged procedure treatment effect & 0.004 & -0.020 & -0.003 & -0.028 \\
Averaged procedure standard error & $ 0.014$ & $ 0.014$ & $ 0.014$ & $ 0.014$ \\
\hline
\hline
Trial & \multicolumn{4}{|c|}{IST3} \\
\hline
Training size & 380 & 760 & 380 & 760\\ 
\hline
Trial treatment effect & \multicolumn{4}{|c|}{0.014} \\
Trial treatment effect standard error & \multicolumn{4}{|c|}{$ 0.034$} \\
\hline
One shot procedure treatment effect & 0.037 & 0.005 & 0.027 & 0.019 \\
One shot procedure standard error & $ 0.034$ & $ 0.034$ & $ 0.034$ & $ 0.034$ \\
Averaged procedure treatment effect & 0.019 & 0.017 & 0.021 & 0.020 \\
Averaged procedure standard error & $ 0.034$ & $ 0.034$ & $ 0.034$ & $ 0.034$ \\
\hline
\end{tabular}
\caption{Treatment effects estimated with n-first control group data augmentation compared to randomised controlled trial.}
\label{tab:nfirst}
\end{table}

\begin{table}[ht]
\center
\begin{tabular}{|l|c|c|c|c|} 
\hline
\multicolumn{5}{|c|}{Sensitivity analysis}\\
\hline
Model & \multicolumn{2}{|c|}{CTGAN} & \multicolumn{2}{|c|}{TVAE} \\
\hline
\hline
Trial & \multicolumn{4}{|c|}{IST} \\
\hline
Training size & 1000 & 5000 & 1000 & 5000 \\ 
\hline
Trial treatment effect & \multicolumn{4}{|c|}{-0.012} \\
\hline
Significative positive treatment effects & 79 & 0 & 76 & 2 \\
Significative negative treatment effects & 464 & 422 & 568 & 562 \\
Incompatible treatment effects & 139 & 0 & 223 & 18 \\
RMSE & 0.018 & 0.006 & 0.022 & 0.010 \\
\hline
\hline
Trial & \multicolumn{4}{|c|}{IST3} \\
\hline
Training size & 380 & 760 & 380 & 760\\ 
\hline
Trial treatment effect & \multicolumn{4}{|c|}{0.014} \\
\hline
Significative positive treatment effects & 208 & 50 & 439 & 88 \\
Significative negative treatment effects & 12 & 0 & 7 & 0 \\
Incompatible treatment effects & 3 & 0 & 28 & 0 \\
RMSE & 0.023 & 0.013 & 0.030 & 0.013 \\
\hline
\end{tabular}
\caption{Characteristics of treatment effects estimated with control-augmented trial data compared to randomised controlled trials.}
\label{tab:sensitivity}
\end{table}

\begin{table}[ht]
\center
\begin{tabular}{|l|c|c|c|c|} 
\hline
Architecture & \multicolumn{4}{c|}{TVAE} \\
\hline
RCT & \multicolumn{2}{c|}{IST} & \multicolumn{2}{c|}{IST3} \\
\hline
Training dataset size & 1000 & 5000 & 380 & 760 \\
\hline
 Compress dimension & 1024 & 2048 & 1024 & 2048 \\
 Decompress dimension & 1024 & 2048 & 2048 & 512 \\
 Embedding dimension & 8 & 4 & 8 & 8 \\
 Batch size & 300 & 100 & 100 & 100 \\
 Loss factor & 4 & 2 & 2 & 4 \\
 Epochs & 1000 & 1000 &  1000 &  1000 \\
 l2-scale & 1e-5 & 1e-5 & 1e-5 & 1e-5 \\
\hline
Architecture & \multicolumn{4}{c|}{CTGAN} \\
\hline
RCT & \multicolumn{2}{c|}{IST} & \multicolumn{2}{c|}{IST3} \\
\hline
Training dataset size & 1000 & 5000 & 380 & 760 \\
\hline
 Generator dimension & 512 & 256 & 512 & 128 \\
 Discriminator dimension & 512 & 1024 & 256 & 1024 \\
 Embedding dimension & 16 & 16 & 16 & 8 \\
 Batch size & 100 & 500 & 500 & 100 \\
 Step & 5 & 5 & 5 & 3 \\
 Epochs & 400 & 400 & 700 & 500 \\ 
 Discriminator decay & 1e-6 & 1e-6 & 1e-6 & 1e-6 \\
 Discriminator learning rate & 2e-5 & 2e-5 & 2e-5 & 2e-5 \\
 Generator decay & 1e-6 & 1e-6 & 1e-6 & 1e-6 \\ 
 Generator learning rate & 2e-5 & 2e-5 & 2e-5 & 2e-5 \\
 Log-frequency & False & False & False & False \\
\hline
\end{tabular}
\caption{Hyperparameters of the $n$-first control group training.}
\label{tab:gs_result_ordered}
\end{table}

\begin{table}[ht]
\center
\begin{tabular}{|l|c|c|c|c|} 
\hline
Architecture & \multicolumn{4}{c|}{TVAE} \\
\hline
RCT & \multicolumn{2}{c|}{IST} & \multicolumn{2}{c|}{IST3} \\
\hline
Training dataset size & 1000 & 5000 & 380 & 760 \\
\hline
 Compress dimension & 2048 & 1024 & 1024 & 2048 \\
 Decompress dimension & 1024 & 1024 & 1024 & 2048 \\
 Embedding dimension & 8 & 8 & 8 & 8 \\
 Batch size & 100 & 100 & 100 & 100 \\
 Loss factor & 4 & 2 & 4 & 2 \\
 Epochs & 500 & 500 &  500 &  500 \\
 l2-scale & 1e-5 & 1e-5 & 1e-5 & 1e-5 \\
\hline
Architecture & \multicolumn{4}{c|}{CTGAN} \\
\hline
RCT & \multicolumn{2}{c|}{IST} & \multicolumn{2}{c|}{IST3} \\
\hline
Training dataset size & 1000 & 5000 & 380 & 760 \\
\hline
 Generator dimension & 512 & 128 & 128 & 128 \\
 Discriminator dimension & 1024 & 256 & 512 & 256 \\
 Embedding dimension & 8 & 16 & 8 & 16 \\
 Batch size & 100 & 100 & 100 & 100 \\
 Step & 5 & 5 & 5 & 5 \\
 Epochs & 400 & 220 & 700 & 500 \\ 
 Discriminator decay & 1e-6 & 1e-6 & 1e-6 & 1e-6 \\
 Discriminator learning rate & 2e-5 & 2e-5 & 2e-5 & 2e-5 \\
 Generator decay & 1e-6 & 1e-6 & 1e-6 & 1e-6 \\ 
 Generator learning rate & 2e-5 & 2e-5 & 2e-5 & 2e-5 \\
 Log-frequency & False & False & False & False \\
\hline
\end{tabular}
\caption{Hyperparameters of the sensitivity analysis training.}
\label{tab:gs_result_sensitivity}
\end{table}

\begin{table}[ht]
\center
\begin{tabular}{|l|cccc|} 
\hline
 \multicolumn{5}{|c|}{TVAE} \\
\hline
 Compress dimension & 256 & 512 & 1024 & 2048 \\
 Decompress dimension & 256 & 512 & 1024 & 2048 \\
 Embedding dimension & 4 & 8 &  &  \\
 Batch size & 100 & 300 &  &  \\
 Loss factor & 2 & 4 &  &  \\
 Epochs & 500 & & & \\
 l2-scale & 1e-5 & & & \\
\hline
 \multicolumn{5}{|c|}{CTGAN} \\
\hline
 Generator dimension & 128 & 256 & 512 & \\
 Discriminator dimension & 256 & 512 & 1024 & \\
 Embedding dimension & 4 & 8 & 16 &  \\
 Batch size & 100 & 500 & 700 &  \\ 
 Step & 1 & 3 & 5 &  \\
 Epochs & 1000 & & & \\ 
 Discriminator decay & 1e-6 & & & \\
 Discriminator learning rate & 2e-5 & & & \\
 Generator decay & 1e-6 & & & \\ 
 Generator learning rate & 2e-5 & & & \\
 Log-frequency & False & & & \\ 
\hline
\end{tabular}
\caption{Hyperparameters considerated in the gridsearches}
\label{tab:gs}
\end{table}

\newpage 
\end{document}